\documentclass[conference]{IEEEtran}

\usepackage{amsmath}
\usepackage{amsfonts}
\usepackage{amssymb}
\usepackage{braket}
\usepackage{pst-node}
\usepackage{tikz-cd} 
\usepackage{setspace}
\usepackage{longtable}
\usepackage{float}
\usepackage{tabstackengine}
\usepackage{hyperref}
\setstackEOL{\cr}

\begin{document}
\title{Simulated Annealing for JPEG Quantization}
\author{
\IEEEauthorblockN{Max Hopkins}
\IEEEauthorblockA{Faculty of Arts and Sciences\\
Harvard University\\
Cambridge, Massachusetts 02138\\
nhopkins@college.harvard.edu}
\and
\IEEEauthorblockN{Michael Mitzenmacher}
\IEEEauthorblockA{Faculty of Arts and Sciences\\
Harvard University\\
Cambridge, Massachusetts 02138\\
michaelm@eecs.harvard.edu}
\and
\IEEEauthorblockN{Sebastian Wagner-Carena}
\IEEEauthorblockA{Faculty of Arts and Sciences\\
Harvard University\\
Cambridge, Massachusetts 02138\\
swagnercarena@college.harvard.edu}}

\maketitle

\begin{abstract}
JPEG is one of the most widely used image formats, but in some ways remains
surprisingly unoptimized, perhaps because some natural optimizations
would go outside the standard that defines JPEG.  We show how to
improve JPEG compression in a standard-compliant, backward-compatible
manner, by finding improved default quantization tables.  We
describe a simulated annealing technique that has allowed us to find
several quantization tables that perform better than the industry
standard, in terms of both compressed size and image fidelity.
Specifically, we derive tables that reduce the FSIM error by over 10\%
while improving compression by over 20\% at quality level 95 in our
tests; we also provide similar results for other quality levels.
While we acknowledge our approach can in some images lead to visible
artifacts under large magnification, we believe use of these
quantization tables, or additional tables that could be found using
our methodology, would significantly reduce JPEG file sizes with
improved overall image quality.
\end{abstract}

\IEEEpeerreviewmaketitle

\section{Introduction}
Since the original development of the JPEG Still Picture Compression
Standard by the Joint Photographic Experts Group in 1992
\cite{JPEGorig}, JPEG has become one of the most widely used image
standards.  Almost a sixth of the space of all web pages is filled by
JPEG images \cite{Guetzli}. JPEG is based on a Discrete Cosine
Transforms (DCT) that shifts the image data into the frequency
domain. JPEG then leverages the fact that lower frequency signals are
more perceivable to the human visual system (HVS) by using a lossy
compression that more aggressively eliminates information related to
the high frequency signal. The key to the exact compression-quality
tradeoff lies in JPEG's quantization table, an eight by eight matrix
that determines to what extent each frequency is compressed. JPEG's
designers emphasized the users' ability to significantly control
visual fidelity, and therefore while a general standard quantization
table has been proposed and adopted in the standard library
\cite{stable}, JPEG implementations allow for the use of personalized
tables. It also includes a quality metric which allows for the tables
to be scaled (the formula for which can be found in the libjpeg code
and documentation \cite{libjpeg}).

JPEG has known deficiencies, but its wide use as an industry standard
has made replacement of JPEG with a new format a seemingly
insurmountable task.  For example, in 2002 Taubman and Marcellin
proposed an improvement on JPEG called JPEG2000 \cite{JPEG200}. While
that paper and later works \cite{JPEG2000study} have shown
that JPEG2000 is an almost universal improvement over JPEG, it is
not widely used and has clearly failed to replace JPEG. 
Our work sets out to improve JPEG compression and image fidelity
without requiring any changes to the JPEG implementations in use
today, specifically by providing improved general quantization tables.
The choice of quantization table is a challenging problem, but
understanding of the problem has improved over time.  When JPEG was
introduced, no reasonable model existed for the HVS.  Little
documentation exists for exactly how the quantization table originally
adopted for JPEG was developed, but it relied largely on testing with
human subjects. One would expect that such a table would have been
determined with a very coarse exploration of the solution space and an
inconsistent measurement of error.  In the early 2000s, the first
promising HVS model was proposed \cite{mssim}, and recent work has
come closer to giving an objective metric closely tied to human
experiments \cite{vif,fsim}.  Such metrics allow us to apply
natural optimization techniques.  

We make use of two of the best current HVS models, FSIM and M-SSIM,
with a particular focus on the first \cite{mssim,vif,fsim}. We have
also incorporated a very recently proposed metric, Butteraugli \cite{Butteraugli}, although its use is
limited to verification due to its computational complexity. With
these error methods we use simulated annealing over a subset of the
Raw Image Dataset (RAISE) \cite{raise} to develop a set of 
quantization tables optimized for both error and compression.  Our
experiments show reductions ranging from 21-50\% in file size depending on the JPEG quality used while improving FSIM error by over 10\% across all qualities. These results are a significant improvement over previous attempts in both compression and image fidelity, and are primarily due to a new, compression focused annealing method.

Our paper proceeds as follows.  We first review the various error metrics
for images.  We then describe our annealing approach, 
focusing on how we aim to improve both compression and image quality.  We
present experimental results demonstrating the improved performance using
our discovered quantization tables, and follow up by diving deeper into our
results, deriving some insight into how our tables improve performance,
and the potential downside of our approach in terms of the possible addition
of compression artifacts.  We conclude describing potential follow-on work.
As a baseline for future work, we provide a small library of quantization tables online. \footnote{\url{http://www.eecs.harvard.edu/~michaelm/SimAnneal/simulated_annealing_for_JPEG_quantization.html}} 
\section{Review of Error Metrics}
The original JPEG quantization tables were created using psychovisual
experiments, basing their results on human perception of the images
\cite{JPEGorig}.  For search-based optimizations methods such as
simulated annealing, requiring human input for the error measurements
is unrealistic, so we employ a machine-computable metric which
reflects human visual perception. Because we are able to compare the
compressed image to the original at any point, we may use
full-reference image quality assessments (FR IQA) that take
advantage of having the original image available. Creating effective
FR IQA algorithms to model the HVS is an active research area.
Over the past decade the more straightforward techniques,
like root mean square error (RMSE) measurements, have been replaced by
more complex and accurate metrics. Sheikh, Sabir, and Bovik provide an analysis
of 10 different modern error metrics, both for color and luminance
only \cite{vif}.  Their analysis suggests that a structural method,
SSIM, is one of the most effective methods in approaching the HSV
standards set by psychovisual databases while also beating other
methods such as VIF (Visual Information Fidelity) in time complexity.
We review RMSE, SSIM, and other error metrics below.  

\subsection{Root Mean Squared Error and Peak Signal-to-Noise}

For many years, IQA methods relied on either Root Mean Squared Error (RMSE) or Peak Signal-to-Noise Ratio (PSNR) methods \cite{RMSE/PNSR}. Part of the reason for this was due to their ease of calculation.  The following are standard equations used in  measuring the ``distance'' between two images:
\begin{align*}
RMSE =& \ \sqrt{\frac{1}{nm}\sum\limits_{i,j=1}^{n,m}(y_{i,j} - \hat{y}_{i,j})^2};\\
PSNR =& \ 20\log_{10}\left(\frac{MAX_I}{RMSE}\right).
\end{align*}

Here $n$ and $m$ are the width and height of the image in pixels respectively, $y_{i,j}$ and $\hat{y}_{i,j}$ are the values of the $(i,j)$th pixel in the two images, and $MAX_I$ is the maximum value for a pixel. While these error methods offer a clear measure of the difference in pixel data, they have little to no correlation with the HVS, and are thus ineffective IQA measurements \cite{RMSE/PNSR}. 
\subsection{SSIM and M-SSIM}
The aptly named Structural Similarity (SSIM) compares
the structural similarity between the original image and the new
compressed image. A seminal paper on the model by Wang et al.,
summarizes the basic concept behind SSIM \cite{mssim}. Essentially
three comparisons are conducted: one on luminescence, one on contrast
(measured through the standard deviation of the image), and one on the
structure (which is a comparison of the images once their mean has
been subtracted and they've been normalized to have standard deviation
1). This basic algorithm was later improved on by Wang et al. and
renamed Multi-scale SSIM, which adds the idea that HSV depends on the scale
at which the image is being viewed, and therefore extending the metric
to multiple resolutions should improve its agreement with
HSV. Mathematically, Multi-scale SSIM only requires that the SSIM
algorithm be run on multiple resolutions and the outputs be
multiplied together by a formula of the form
\begin{align*}
\text{Multi-scale SSIM} &= \prod \left(\text{SSIM}_i \right)^{\gamma_i},
\end{align*}
where each $\gamma_i$ represents a weight for the importance of a particular resolution, the $i$ index iterates over the the different resolutions being sampled, and the product is over the SSIM scores at each resolution. Unsurprisingly, Multi-scale SSIM has been shown to outperform SSIM and has given some of the best results on HSV databases.
\subsection{FSIM}
While several broad comparison papers have pointed to VIF and SSIM as
the leading FR IQA algorithms, a more recent publication has put forth
a new technique, labeled feature similarity (FSIM), and shown that
FSIM outperforms both SSIM and VIF on six major publicly available IQA
databases \cite{fsim}. FSIM takes advantage of two newer strategies in
the IQA community: phase congruency (PC) and gradient magnitude
(GM). PC is based on the idea that we generally perceive features
in locations where the Fourier components are in phase, which is a
refinement over SSIM's simplistic focus on contrast. The GM strategy,
on the other hand, is effectively a measure of contrast, allowing FSIM
to factor in SSIM's basic feature analysis. The final FSIM score is
then the weighted product of the similarity between the GM values and
the similarity between the PC values. Instead of summing over the
similarity of each point, the sum is weighted by the maximum of the two PC
values at the point, which effectively assigns more weight to
high importance regions. 

\subsection{Butteraugli}

Google Research recently introduced a new error method named
Butteraugli \cite{Butteraugli}. While there do not yet appear to be
any published results for the metric, the
specification for Butteraugli claims that the method performs better
than SSIM or PSNR on tests where overall compression is roughly equivalent
to quality 90-95 of the standard JPEG table. While more details need
to be released, Butteraugli uses a heat map
of differences between the original and compressed images for its
computation. Butteraugli also takes significantly longer to calculate on a single
photo than comparable methods such as FSIM, but 
could be more efficient when
calculating the metric on similar versions of a photo;  this is a potentially useful
feature for simulated annealing that we believe should be explored in future work. 

\section{Related Work}

The first attempt to use simulated annealing techniques on JPEG
quantization tables was a 1993 paper by Monro and Sherlock
\cite{MS1993}. Their initial technique worked with all 64 quantization
coefficients and utilized RMSE as their error metric. Due to their
computational limitations, they were only able to anneal on a single
image, Barabara 2. Their results gave an 8\% improvement on error to
the standard table. However, as discussed previously, RMSE has
been shown to be a poor metric for the HVS. Their work also
demonstrated that the high frequency domain
was more important than the quantization by the standard table
assumed. A paper published a year later by Sherlock et
al. \cite{MS1994} proposed a simpler 3 parameter model for annealing
that minimized computational difficulties. The results showed a
similar order of magnitude improvement against the standard table. The
paper also highlighted the ineffectiveness of RMSE by showing that a
one parameter model (a constant value for all quantization values)
gave only marginally worse RMSE error as compared to standard table despite a
significant visual difference. Later work 
by Monro et al. in 1996 \cite{MS1996} reaffirmed their previous work
and expanded the discussion of JPEG optimization using simulated annealing
to varying block sizes. While their approach to modifying block sizes showed
promise, it would require a significant modification of the JPEG
algorithm.

As new statistical methods and models of the HVS arose, the question
of the quantization table was tackled again.  Direct frequency analysis
gave tables with slightly less error, amounting to a around 3\%
improvement on PSNR error \cite{Battiato}. Some work extending on the
contribution by Sherlock et al. was done using genetic algorithms, but
the results gave at best a 10\% improvement on PSNR \cite{Genet}.

More recently, Pattichis and Jiang considered Pareto
optimal points, quantization tables such that no other table with a lower
bit-rate and a higher SSIM score exists. They proposed five annealing techniques to
approximate these points, the best of which offered a sizable
11.68\% improvement on bit-rate over their training set. However, Pareto optimal points differ by photo, and the authors did not report results over a disparate evaluation set, which we would expect to significantly decrease their performance.

Upon testing variants of the five proposed techniques, we adopted a method similar to their suggestion of using an exponential function in frequency to change higher frequency values of the quantization table more often. Our variant differs slightly in how it selects these higher frequency values, and in addition changes ten values per step, rather than the one suggested by Pattichis. Further, due to the shape of the error solution space, instead of directly attempting to approximate Pareto
optimal points, our annealing variant focuses on compression
maximization with a temperature function that rewards lower
error. Lastly, the error metric used in their paper, SSIM, while better
than those in the early works, is still out-dated when compared to
M-SSIM and FSIM.

Within the last few months, Google Research Europe launched the new
JPEG Encoder ``Google Guetzli'' \cite{Guetzli}.  While the methods
used in their paper differ significantly from ours, we note the
techniques are mostly orthogonal, and our approach to generating
tables could be combined with Guetzli's innovative methods. Rather than using globally improved tables, Guetzli
runs a brief annealing process on every photo it compresses, using the
Butteraugli method to maximize compression while maintaining image
fidelity. However, most of the improvement from Guetzli comes from an
improvement not to the quantization table, but to the DCT coefficients
themselves. Guetzli uses RLE-encoding on the zeroes in the DCT
coefficients that causes additional adjacent zeroes to have a very low
compression cost. Thus their main compression method consists of
identifying DCT coefficients to set to zero without greatly altering
the Butteraugli error score of the resulting image. When combining the
annealing on Butteraugli with this method, Google produced some
impressive results, claiming to be able to improve image compression
by 29-45\% without losing image fidelity. Due to both methods being specific to each image, this fundamentally differs from our goal of creating a generally optimized quantization table. We believe that
combining our initial general tables with these methods could boost
the results of both papers;  we leave this exploration as future work.
\section{Annealing}
\subsection{Standard Table}
While many specialized quantization tables exist for photography applications, any of the numerous applications that directly employ libjpeg use the luminance table adopted in their standard library \cite{stable}. We call this table the standard quantization table, shown below:
\begin{equation*}
\setstacktabbedgap{4pt}
\text{Standard} = 
\bracketMatrixstack{
16 & 11 & 10 & 16 & 24 & 40 & 51 & 61 \cr 
12 & 12 & 14 & 19 & 26 & 58 & 60 & 55 \cr  
14 & 13 & 16 & 24 & 40 & 57 & 69 & 56 \cr  
14 & 17 & 22 & 29 & 51 & 87 & 80 & 62 \cr 
18 & 22 & 37 & 56 & 68 & 109 & 103 & 77 \cr  
24 & 35 & 55 & 64 & 81 & 104 & 113 & 92 \cr  
49 & 64 & 78 & 87 & 103 & 121 & 120 & 101 \cr  
72 & 92 & 95 & 98 & 112 & 100 & 103 & 99 
}
\end{equation*}

Our goal is to propose a new baseline table that offers improvements to the standard table over standard image sets.
\subsection{Annealing variants}
We recall the basic framework of simulated annealing.  At each step, we have a solution;  in our case, a solution is a quantization table.  We randomly locally perturb the current solution to obtain a neighbor of that solution, and update our current solution to be that neighbor with some probability, depending on a scoring function.  In particular, we may move to a solution that is worse than the current solution with a probability that depends on the change in score and the number of steps the process has run, according to what is commonly referred to as a temperature function.  

We discuss the temperature function in detail below, but describe some of our other decisions.  After experimenting with various methods of generating neighboring candidates in our annealing, we settled on selecting ten values (with replacement) and randomly altering these by $\pm 1$ chosen with probability inversely scaling with the importance of that frequency to the HVS. We roughly approximate a frequency's importance by its value in the standard table (a smaller value indicates higher importance). Selecting ten values at once allows our annealing to move very quickly through the solution space--necessary due to its gargantuan size. However, such a method makes it easier for poor changes to be packaged with positive changes in a single move. For example, if 9 of our 10 changes results in worse compression and error but one gives large improvements on both the step could be considered a net positive and taken. This could have implications for our ability to reach an optimum, but since (as we describe later) we are running 400 annealing processes in parallel we believe the approach is still appropriate.

We also compare annealing processes that minimize error with those that maximize compression. In the minimization version of our annealing, steps with lower error are always accepted unless they reduce compression by more than one percent of the standard table's compression. Steps with higher error are accepted with some probability given by the temperature function. Similarly, in the maximization version, steps with higher compression are always accepted unless they increase error by more than one percent, and steps with lower compression are accepted with some probability.
\subsection{Probability and Temperature Functions}

\begin{figure} [h!]
\includegraphics[scale=.7]{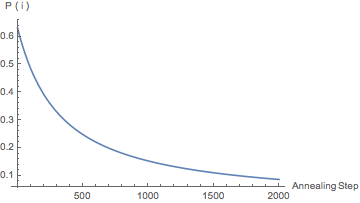}
\caption{Our acceptance probability for a solution with 1\% larger file size and 1\% smaller error value than the baseline}
\label{fig:annealingfunction}
\end{figure}

We now describe our temperature function and how it relates to the probability of updating our current solution at each step. Let $C^*$ and $E^*$ be the compressed file size and error of the baseline solution, where the baseline is the most recently accepted solution. To be more specific, $C^*$ is the total file size of the training set (detailed in the following section) compressed with the current solution, and $E^*$ is the sum of the errors of the same compressed images. For the results presented in this paper, the training set consists of ten images and the error is measured using FSIM. We let $S^* = (E^* \times C^*)^{20}$ be the ``score'' for the current solution.  Similarly, let $C_i$ and $E_i$ be the compression and error of the proposed solution over the ten image training set of images at the $i$th step, and let $S_i = (E_i \times C_i)^{20}$.  Finally, let the temperature $T(i)$ be $200/(200+i)$.  Then the probability $P(i)$ we move to the new solution at the $i$th step is given by
\[
P(i) = \begin{cases}
1 & \text{if} \ C_i < C^*; \\
\min\left[1-e^{S_i/(T(i)S^*)},1\right] & \text{otherwise.}
\end{cases}
\]
In the above equation we have assumed that we are optimizing for compression, but the same equation could be used to optimize for error if we substitute $E_i < E^*$ for $C_i < C^*$.
The constants in the temperature and score functions are manually tuned such that the probability function approaches zero near the mid thousands for annealing candidates with small fluctuations in error and compression. We experimentally found the mid thousands to be an ideal cutoff, with a larger or smaller value yielding poorer results. In addition, the exponent in the score functions are chosen so that steps with large reductions in error are still likely to be kept even late in the annealing. For example, a step that yields even a small increase in size along with a 1\% decrease in error has a 10\% chance of being taken on the $2000^{th}$ step, while one with a 10\% decrease in error has a 45\% chance of being taken. Again, the exact choice of sensitivity to reductions in error was chosen from experimental results. Figure \ref{fig:annealingfunction} displays our acceptance probability $P(i)$ for a typical annealing candidate with 1\% larger file size and 1\% smaller error value than the baseline.
\subsection{Implementation and Methods}
\begin{figure}[h!]
\includegraphics[scale=.4]{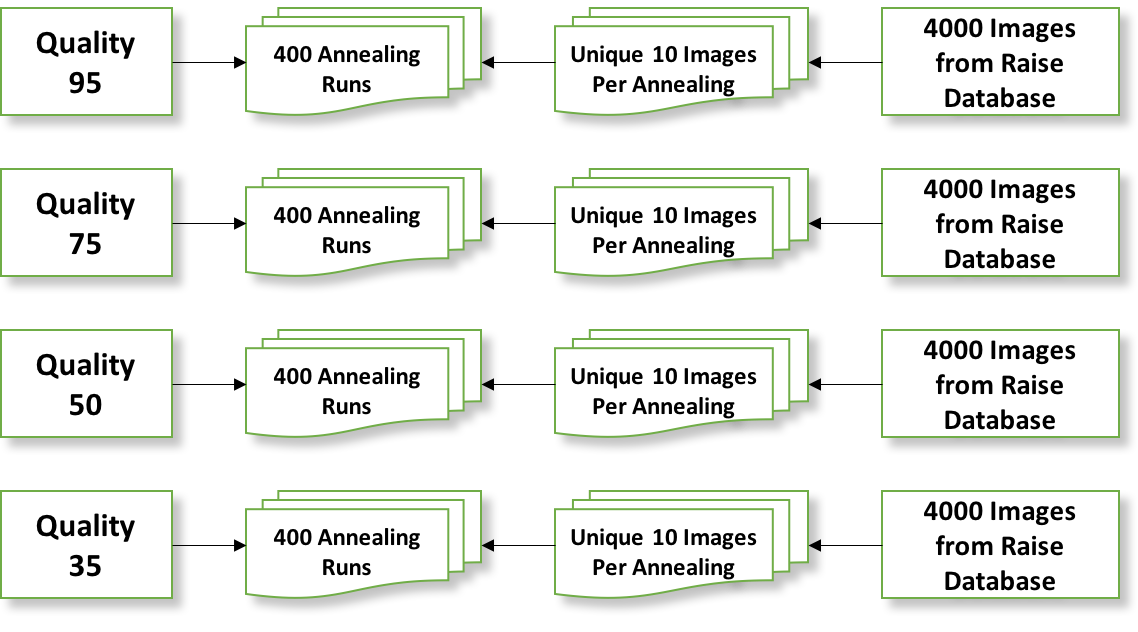}
\centering
\caption{Above is a basic schematic of our implementation. For each quality we anneal on, we have 400 runs. Each of the 400 runs has its own unique subset of 10 images that it runs on, pulled from a set of 4000 training images from RAISE.}
\label{fig:implm}
\end{figure}

Our implementation of annealing uses the cjpeg command line tool from the libjpeg library at each step to calculate the compression ratio $\left (\frac{\sum \text{New Image Size}}{\sum\text{Standard Table Image Size}}\right )$ of the candidate quantization table over a set of images \cite{libjpeg} at a given cjpeg quality parameter. Further, we use an implementation of FSIM due to L. Zhang \cite{fsim} at each step to calculate error. We strictly use FSIM both due to its greater speed, allowing far more steps than methods such as Butteraugli, and the results in \cite{fsim} suggesting it better models the HVS than then existing alternatives. Lastly, we used Harvard's computing cluster Odyssey to run and parallelize annealing and data analysis. The main challenge in deciding specifications for the annealing technique came in the trade-off between the generality of the quantization tables produced by each annealing run, and the total number of steps possible given the time per step. Calculating the average compression and error values over a larger database of images gives greater generality but lowers the number of steps through which the annealing may explore the solution space. In the end, despite being able to parallelize the work over Odyssey, calculating compression and error for each candidate table in the annealing over any significant number of images was computationally inviable due to the start-up time of Matlab, the time to receive a node on the cluster, and node failure within the cluster. Thus, as is depicted in Figure \ref{fig:implm}, we ran 4 groups of 400 separate annealing processes in parallel at qualities 95, 75, 50, and 35. Each process was trained on 10 distinct images after partitioning the database of 4000 RAISE images \cite{raise} into 400 groups (at random), which we will refer to as the training set.

Running over 10 photos kept a reasonable level of generality and allowed for over 3000 steps of annealing over a 7 day period (with the actual values varying by node). This amounted to about 3.36 minutes per step of annealing. Further, since our annealing method probabilistically selects changes, running 400 separate processes better explores the enormous QT solution space.
\subsection{Compression vs. Error Optimization}
\begin{figure} [h!]
\includegraphics[scale=.21]{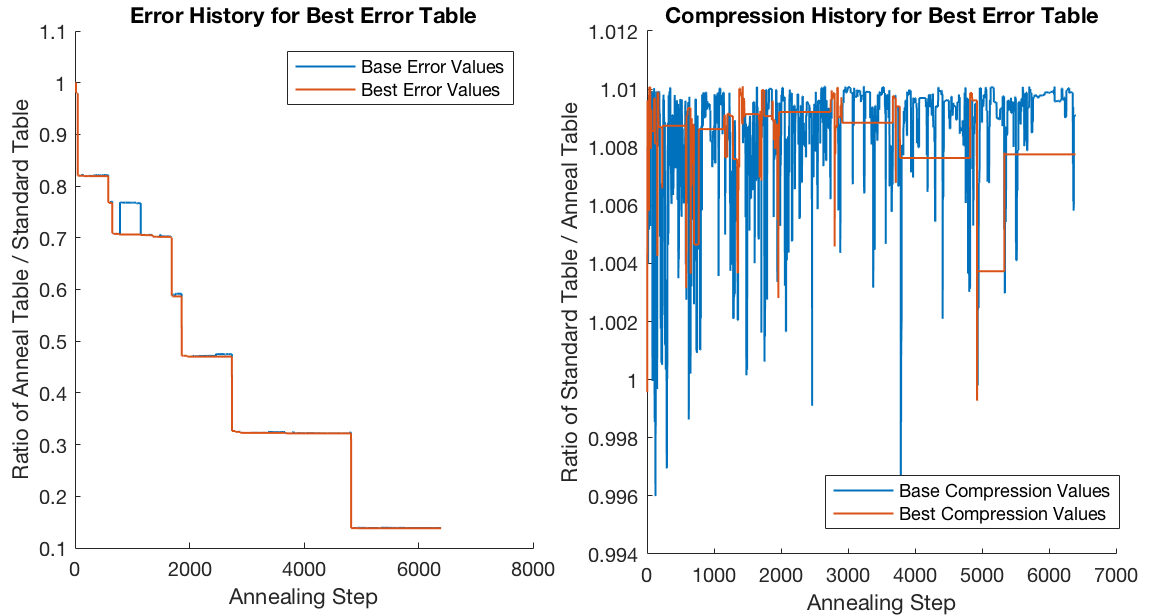}
\centering
\caption{These two graphs show the error and compression ratio throughout the annealing run that generated the best (lowest) final error ratio. The blue line in both the left and right hand graph is the value of the baseline table at that particular step. The orange line is the value of the ``best'' table seen so far, where best in this case means the lowest error. The error history on the left hand side resembles a step function, with each step in the annealing either having no real effect on the error or making a drastic change. The right hand side has no real discernible trend.}
\label{fig:best_error_hist}
\end{figure}

\begin{figure} [h!]
\includegraphics[scale=.21]{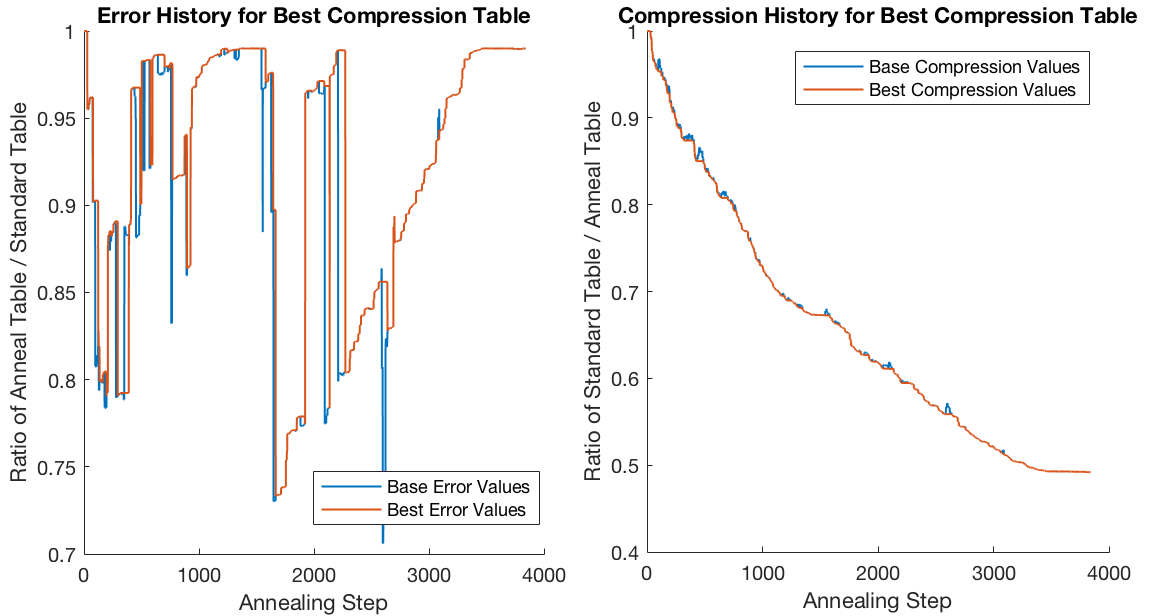}
\centering
\caption{These two graphs show the error and compression ratio throughout the annealing run that generated the best (lowest) final compression ratio. The blue line in both the left and right hand graph is the value of the baseline table at that particular step. The orange line is the value of the ``best'' table seen so far, where best in this case means the lowest compression ratio. The error history on the left hand side still has some clear steps and plateaus, but also has periods of gradual change. The compression history on the right shows a near constant gradual decline.}
\label{fig:best_comp_hist}
\end{figure}

Initially, we set out with the objective of generating three types of tables in our annealing: tables that minimize error, tables that maximize compression, and tables that improve both. However, we quickly came to realize that annealing on error produces far less favorable results than compression, especially if our goal is to find tables that simultaneously improve both metrics. While error annealing would deliver the tables with the lowest error, as one would expect, it would always come coupled to a substantial decrease in compression. The same could not be said for our compression runs, which consistently delivered sizable improvements to both metrics in the same period of annealing. A side by side comparison of the annealing history of our best performing tables for both metrics helps explain this asymmetric result. We plot the annealing graphs for the best error optimizing table in Figure \ref{fig:best_error_hist} and best compression optimizing table in Figure \ref{fig:best_comp_hist} over our 4000 picture training set, where error is measured by the ratio $\left (\frac{\sum (1 \ - \ \text{New FSIM Score})}{\sum (1 \ - \ \text{Standard Table FSIM Score})} \right )$ over a set of ten images at quality parameter 75. In both figures the FSIM error metric shows a clear pattern of plateaus and sudden jumps. While the error optimizing run manages to reach a much lower final error ratio, both annealing runs find tables with noticeably improved error. However, the annealing of the compression optimized table in Figure \ref{fig:best_comp_hist} moves smoothly through the compression space, whereas the error optimized table in Figure \ref{fig:best_error_hist} moves through the compression space almost randomly, never producing a meaningful improvement. As we discussed previously, both annealing runs use the same randomized policy to choose the next table to test. Therefore, it is solely the decision to prioritize compression optimization that yields the more balanced results of Figure \ref{fig:best_comp_hist}.


There are a few reasons for the discrepancy between how the processes move over the space of possible tables. The first is that FSIM, despite being one of the best metrics in the current literature for modeling the HVS, has difficulty modeling small changes in error. Annealing on a metric that resembles more a step function than a continuous function gives poor results. The second is that there seem to be, at least on the surface, some inefficiencies in the standard table we are remedying. It is natural that, generally, when our compression increases so does our error, but there are steps where error will decrease drastically while compression is only minimally changed. Similarly, there are steps where compression increases and error is weakly affected. Therefore, due to the quantized nature of the error, as we move along our compression space we stumble upon steps that make huge improvements to error but barely change compression. In these cases our annealing function accepts with high probability. Then our compression will continue to improve while error only occasionally increases. The net effect is that eventually we find ourselves at better compression and error. Finding this same improvement while optimizing in the error space is effectively impossible since compression changes by fractions of a percent from one annealing step to the other. Our temperature function is intentionally insensitive to such small changes, as we found that optimizing for both metrics simultaneously produced incredibly poor results.
\section{Results}
\begin{figure*}[h!]
\includegraphics[scale=.54]{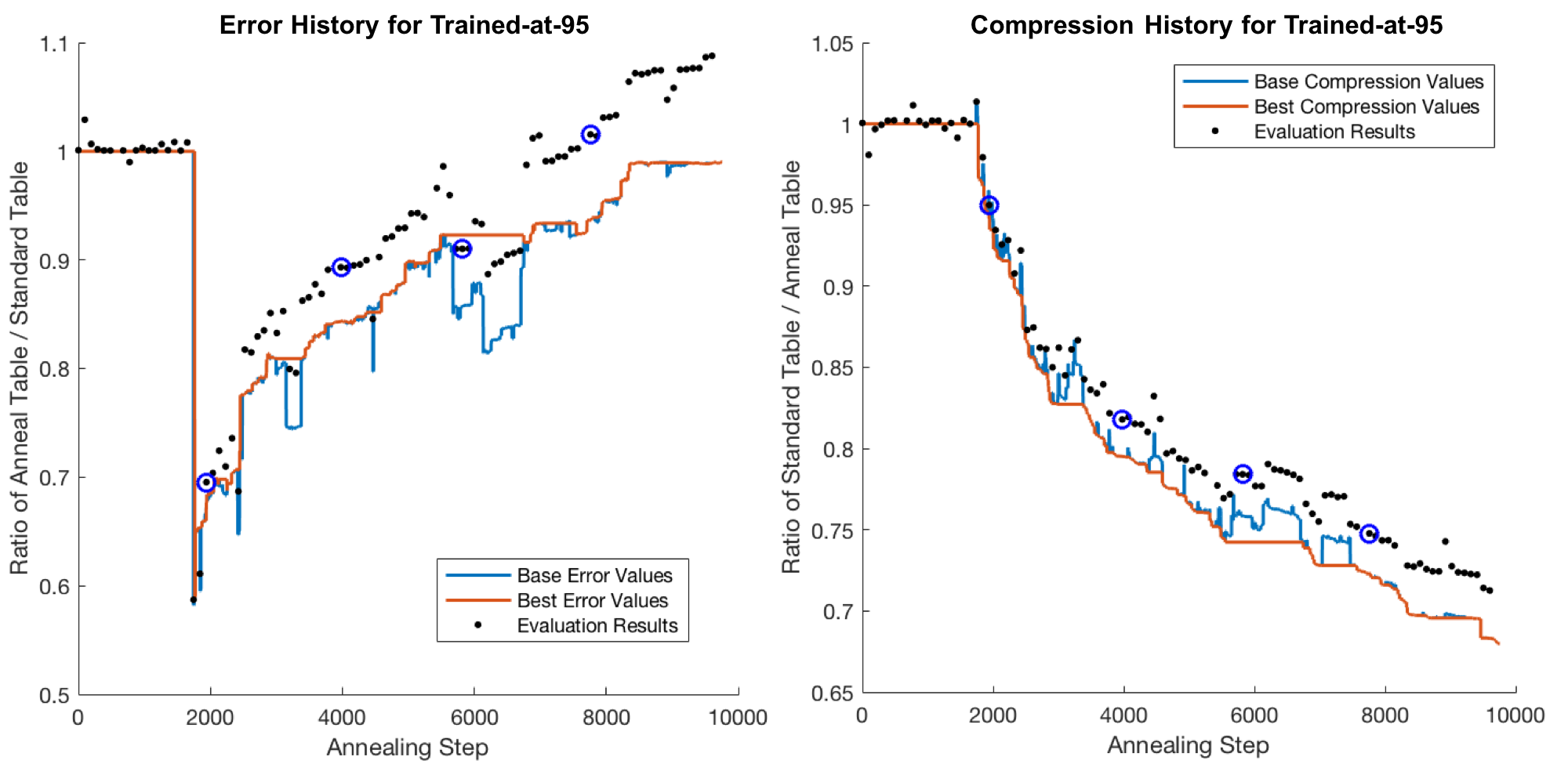}
\caption{These two graphs show the error and compression ratio throughout the annealing run that generated the best (lowest) final compression ratio trained at quality 95. The blue line in both the left and right hand graph is the value of the table being tested at that particular step, and the orange line is the value of the table with the lowest compression seen up to that step in the annealing. The black dots correspond to the error and compression ratios over our evaluation set for the 100 tables we chose to sample from this history. A few of these points have been removed due to hanging processes on the Odyssey cluster. The gap between the black dots and the blue line represents the gap between performance on our training and evaluation sets. The black dots circled in blue are the tables used for the visual comparison in Figures \ref{fig:95compiled_orig} and \ref{fig:95compiled_crop}}
\label{fig:95_hist}
\centering
\end{figure*}
For each of the annealing qualities (95,75,50,35), we compared all the runs based on the best table found with regard to compression, selected the five runs that performed the best, and calculated their error and compression ratios across 200 images from the RAISE database distinct from the training set \cite{raise}. We refer to this set as our evaluation set.
For each quality, we then selected the single table with the best results, based on a trade-off of error and compression ratios. However, we were also interested in sampling the evolution of the best table throughout the annealing process. Therefore, we subsampled 100 tables from the history of our best table and ran them over the evaluation set. The final tables presented in Figure \ref{fig:heatmap} are those from this subsample of 100 that achieve the highest compression while still improving FSIM error by over $10\%$ over the evaluation set. The improvements in compression for these four tables range from $20-50\%$ depending on the training quality. In the next two sections, we walk through the process described above in further detail, provide image comparisons to corroborate the final tables' improved FSIM scores, and give a brief summary of the properties that seem to allow these tables to outperform the standard table.

\subsection{Annealing History}
\begin{figure*}[h!]
\centering
\includegraphics[scale=.035]{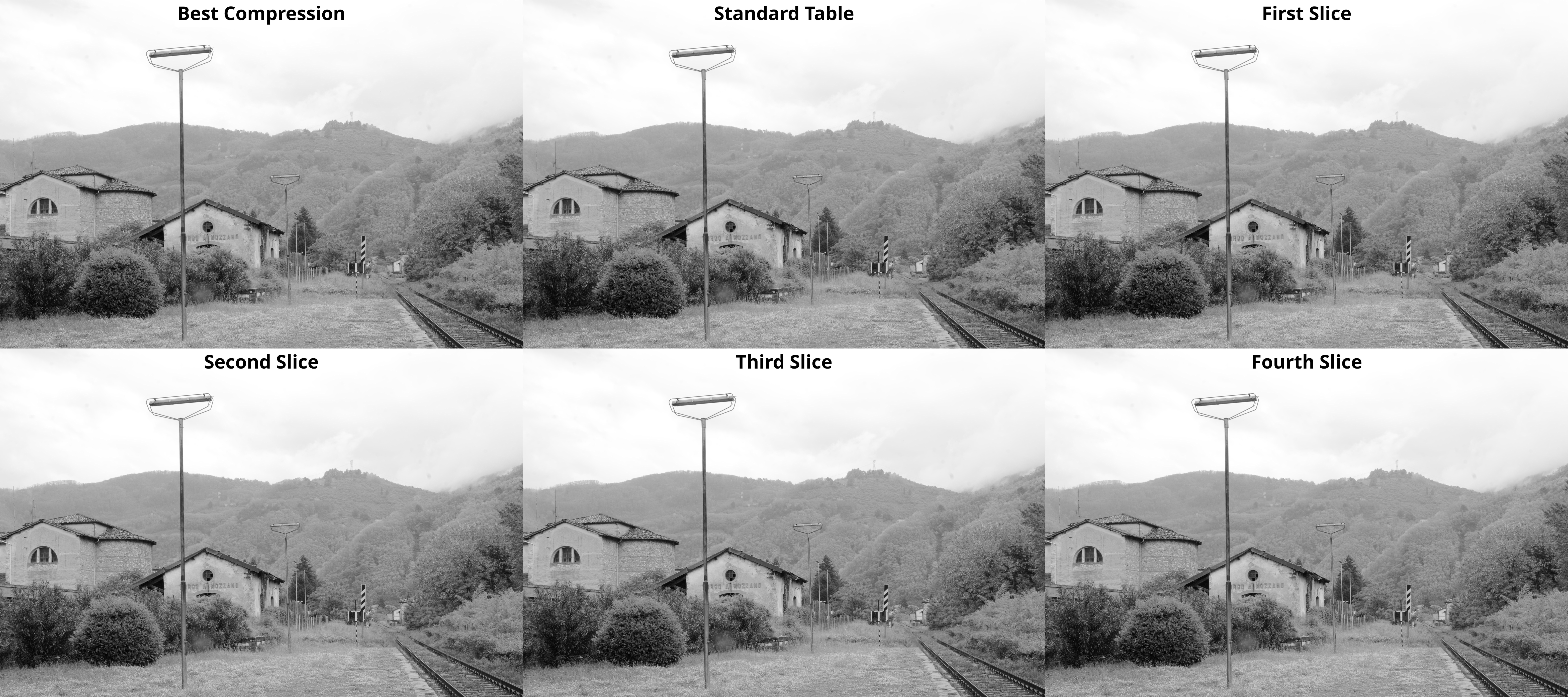}
\caption{These six unmagnified images give a visual comparison of the image fidelity throughout our quality 95 annealing process shown in Figure \ref{fig:95_hist}. The upper left is the best compression from the entire history, followed by the standard table, and then four tables in chronological order circled in Figure \ref{fig:95_hist}. There is no obvious visual difference between any of these photos. This version is cropped for arXiv space limitations, for full image visit: \url{http://www.eecs.harvard.edu/~michaelm/SimAnneal/PAPER/simulated-annealing-jpeg.pdf}}
\label{fig:95compiled_orig}
\end{figure*}
\begin{figure*}[h!]
\centering
\includegraphics[scale=.25]{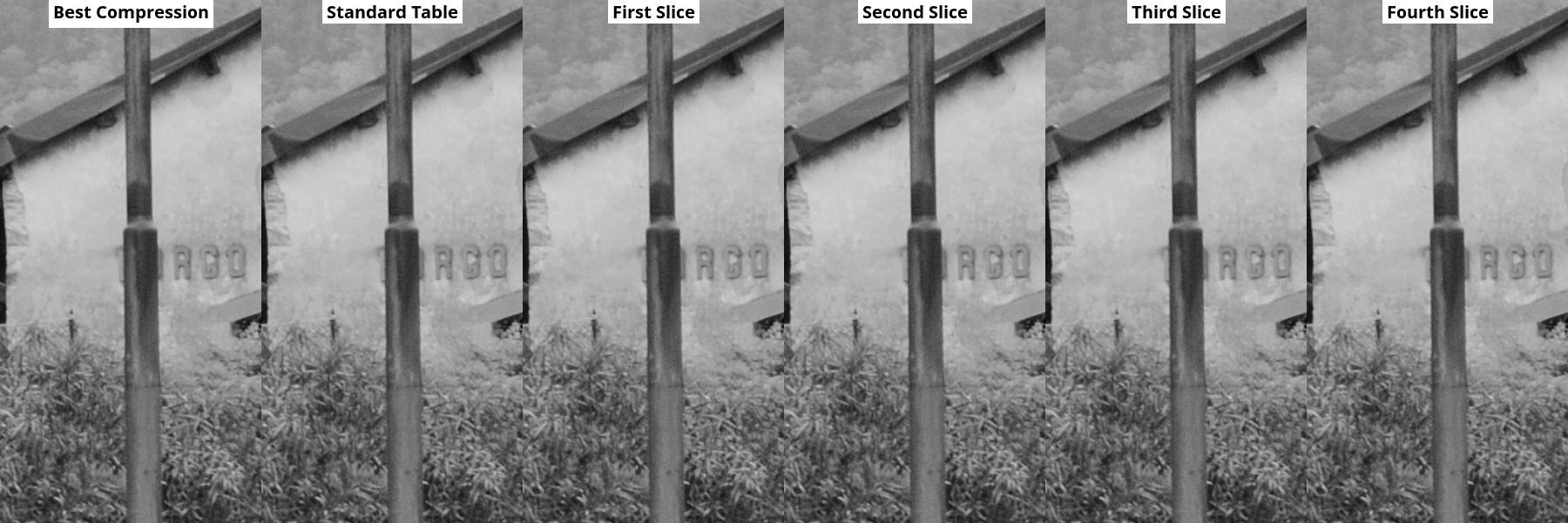}
\caption{These six images are magnified and cropped versions of those in Figure \ref{fig:95compiled_orig} and give another visual comparison of the image fidelity throughout our quality 95 annealing process. Even with such magnification there is no obvious visual difference between any of these photos.}
\label{fig:95compiled_crop}
\end{figure*}
As we discussed previously, in order to better understand our annealing process we took the best performing table for each quality and calculated compression and error ratios for 100 tables from its annealing history, each one one-hundredth of the way through, across our evaluation set. The results for all four qualities can be found in the appendix in Figures 12, 13, 14, and 15 \footnote{The appendix for the arXiv version of this paper does not include these figures due to arXiv size restrictions. For the full appendix please see \url{http://www.eecs.harvard.edu/~michaelm/SimAnneal/PAPER/simulated-annealing-jpeg.pdf}}, but here we will specifically discuss the results for our quality 95 run. Figure \ref{fig:95_hist} shows the full annealing history for the best table trained at quality 95. As with the graphs we presented in the previous section, the left hand graph shows the history of the error ratio $\left (\frac{\sum (1 \ - \ \text{New FSIM Score})}{\sum (1 \ - \ \text{Standard Table FSIM Score})} \right )$ and the right hand side the history of the compression ratio $\left (\frac{\sum \text{New Image Size}}{\sum \text{Standard Table Image Size}}\right )$, both over the training set with quality parameter 95, and where in both cases a lower value is an improvement. The blue line in both graphs is the value of the most recent table that was kept (the baseline table), and the orange line is the value for the table with the lowest compression ratio the annealing algorithm has seen overall. Recall that the most recently accepted table is not necessarily the best table seen so far. Both the blue and the orange line show values on the original 10 images used to train the annealing. The black dots correspond to the 100 tables we sampled from the annealing history, and they represent the value of the current table at that step on the 200 image evaluation set. Therefore, the gap between the black dots and the blue line represents the difference between the performance of our tables on the training and evaluation sets.

There are a couple of interesting trends to note. The first is that the graphs show that despite only training on 10 images, most of our results carry over to the evaluation data. The shape of our error and compression graphs is mirrored closely by evaluation results, with the error deviating by a range of 1.84\% to 6.35\% and compression by .76\% to 3.27\% between the 2000th and 6000th step. The gap grows in the later annealing steps, extending to between and 6.17\% to 9.81\% for error and 1.94\% to 4.41\% for compression from the 6000th step to the end of the annealing. This implies that our annealing does begin to exploit specific features of the images in its 10 image training set, but this is not entirely unexpected, as inevitably the annealing will choose a table that only makes substantial improvements for the subset of 10 training images. What is reassuring is that there is no step where improvements on the 10 images used for training do not translate to improvements on the completely independent 200 image evaluation set. Interestingly, for quality 50, the final 3000 steps perform better on the evaluation set than the training set for both metrics, and for quality 75 the error on the evaluation set is consistently lower than on the training set (the graphs for these qualities can be found in the appendix in Figures 12 - 14). This suggests that in part the gap in results between evaluation and training data is also due to how easily our ten image subset can be compressed compared to the mean, which we assume the 200 image evaluation set better captures. The comparison of our results on the training and evaluation sets strongly suggests that our tables generalize well. While we were restricted by computation limitations, this gap could also be reduced by expanding the training set beyond our choice of ten images. 

\begin{figure*}[t!]
\centering
\includegraphics[scale=.25]{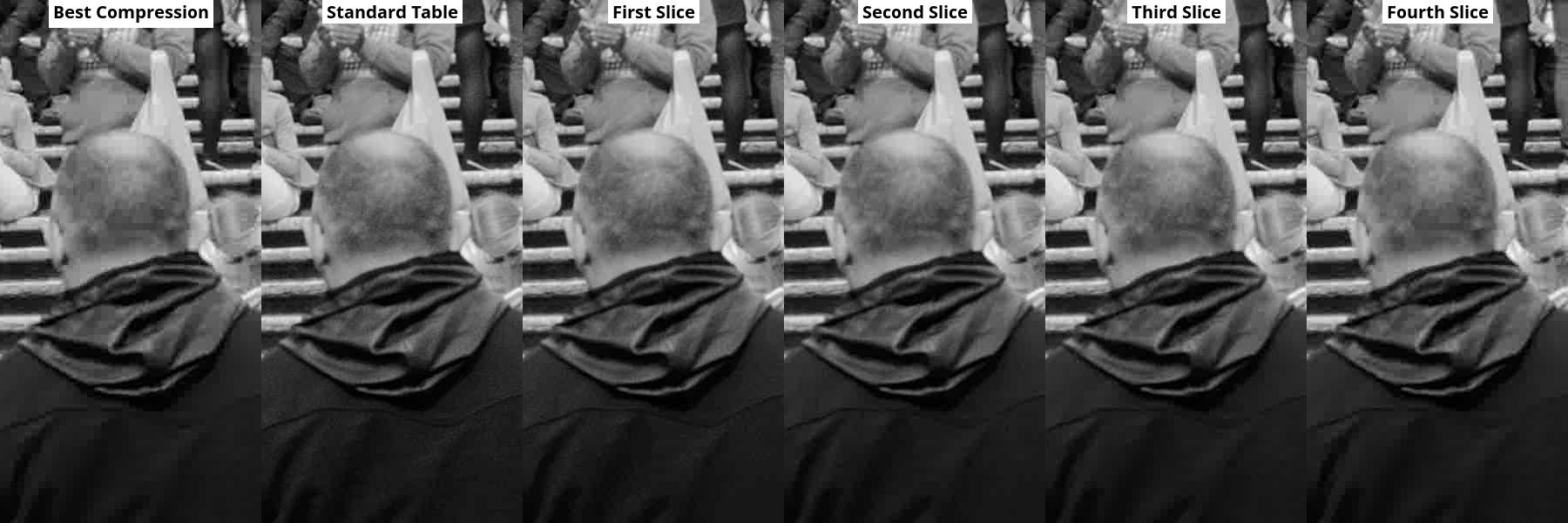}
\caption{These six images are magnified and cropped versions of the original pictures found in Figure 12. Going from left to right, they were generated by the table with highest compression, the standard table, and the four tables circled in the graph of 12. They show some signs of pixelation in the most compressed images. In particular, this pixelation can be seen in the comparison between the picture labeled 'standard table' and the one labeled 'best compression'.}
\label{fig:35compiled_crop}
\end{figure*}

In Figures \ref{fig:95compiled_orig} and \ref{fig:95compiled_crop} we present a side by side comparison of the compression of an image using different tables from the annealing history presented in Figure \ref{fig:95_hist}. We chose 6 tables for our visual comparison, the table with the best compression that our annealing found, the standard table, and four tables that divide the compression history into fourths. These four tables occur at steps 1940, 3977, 5820, and 7760. Their evaluation results are circled in blue in Figure \ref{fig:95_hist}. Note that while we focus on the quality 95 visual comparison here, the same visual comparison for all four qualities can be found in Figures 12 - 15 in the appendix. To our best judgment the unmagnified side by side images in Figure \ref{fig:95compiled_orig} are not distinguishable, as are their significantly magnified counterparts in Figure \ref{fig:95compiled_crop}. We can make the same claim of the unmagnified images for all three of the other qualities. However, in Figure \ref{fig:35compiled_crop}, which shows a side by side comparison of the magnified images for quality 35, there is a small drop in fidelity. If you look closely, you may notice some additional pixelation on the images labeled Best Compression and Fourth Slice. This is not shocking, given that a fifty percent increase in compression requires the loss of some information. However, there are two factors we believe minimizes the relevance of this pixelation.

The first is that it is noticeable only upon significant magnification of the image, and it does not bring with it any effects on the unmagnified image (for example, as we will show in Final Results section, the same distinction does not hold for compression through quality scaling for the standard table). Also the effects are only noticeable for the later tables, where we have previously discussed our annealing may begin to exploit its training set in ways that do not extrapolate to a larger image set. Therefore, for our final results we select tables earlier in the annealing where the error on the evaluation set gives at least a 10\% improvement. However, it is possible that part of the reason we see this pixelation is because our annealing has found a way to exploit FSIM. As we noted in the error section, FSIM is not a perfect model of the HVS, and therefore it should be possible to decrease FSIM error in ways that do not correspond to HVS perceived error. In fact, despite improved performance with FSIM, our tables perform slightly worse with regard to the Butteraugli metric. While the compression values of our final table trained at quality 95 fall somewhere just below the standard table with quality parameter set to 93, its Butteraugli score is closer to the standard at quality 92. It may be that Butteraugli is picking up on the pixelation, and it is thus possible that a combination of FSIM and Butteraugli may be able to address this issue. We will discuss this possibility in our Future Work section.
\subsection{Final Tables}
\begin{center}
\begin{tabular}{|c | c | c |}
\hline
\multicolumn{3}{|c|}{Top Optimization Values} \\
\hline
\textbf{Table Name} & \textbf{Error Ratio} & \textbf{Compression Ratio}  \\
\hline
Trained-at-95 & .8963 & .7872 \\
\hline
Trained-at-75 & .8619 & .5890 \\
\hline
Trained-at-50 & .8873 & .4984 \\
\hline
Trained-at-35 & .8394 & .5737 \\
\hline
\end{tabular}
\end{center}

\begin{figure*}[h!]
\centering
\includegraphics[scale=.8]{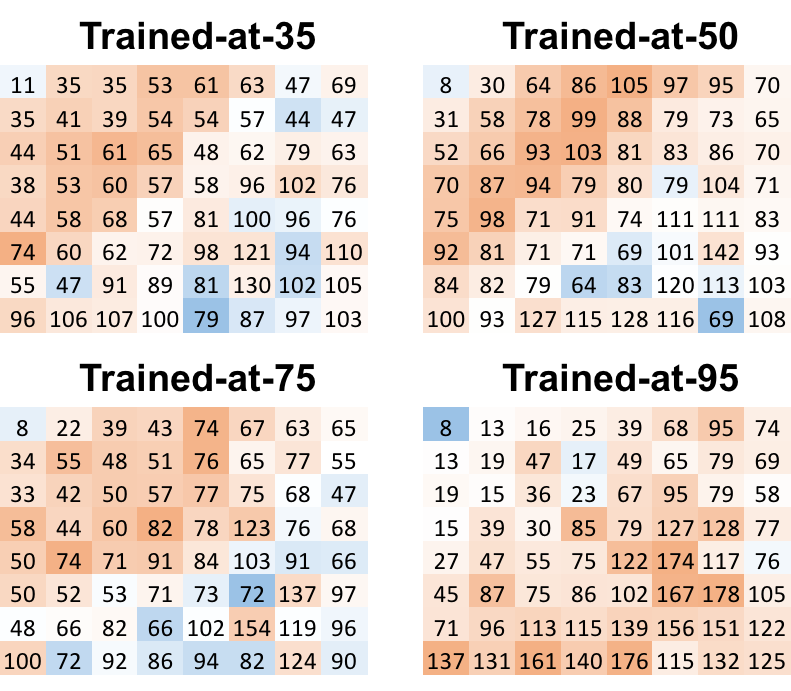}
\caption{Heat map of final quantization tables. Red tints denote an increase in value, blue a decrease. The shade of the tints correlates to the magnitude of the change.}
\label{fig:heatmap}
\end{figure*}

We have chosen one table trained at each quality in the aforementioned manner to present as the best table. The error and compression ratios are calculated as above across our 200 image evaluation set at the same quality parameter at which they were trained. Note that along with those specifications, we selected tables whose FSIM error is improved by at least 10\% over the standard table on this evaluation set. Figure \ref{fig:heatmap} shows these tables with respect to the standard table, where red values have been increased, blue values have decreased, and intensity corresponds to the relative magnitude of change normalized by the maximum and minimum changes in the table.

\subsection{Visual Improvement}
\begin{figure}[h!]
\centering
\includegraphics[trim={20cm 22cm 0 0}, clip ,scale=.055]{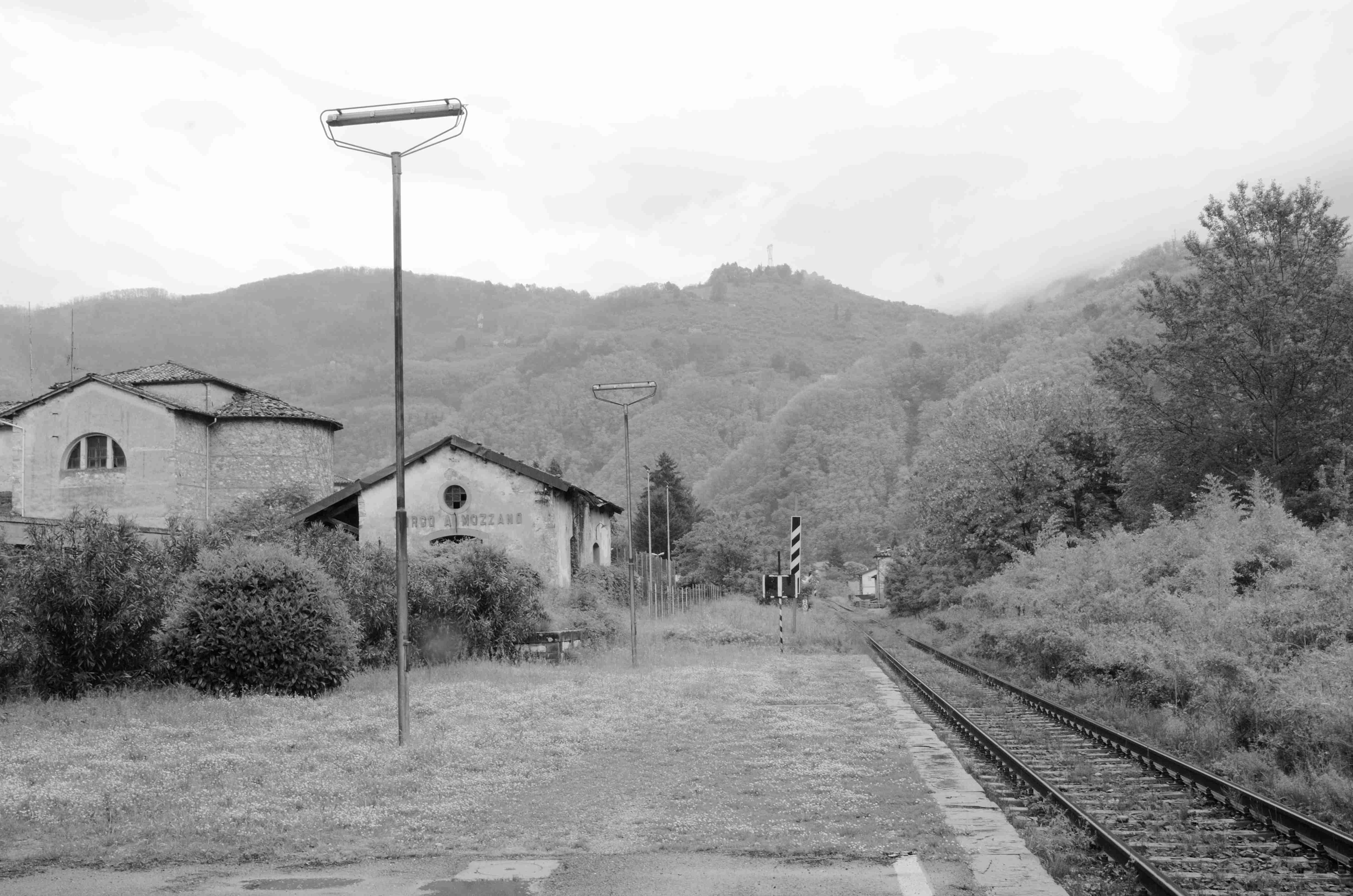}
\caption{Image compressed with low quality parameter by the standard table. The sky shows clear gradation}
\label{fig:low_quality}
\end{figure}

\begin{figure}[h!]
\centering
\includegraphics[trim={20cm 22cm 0 0}, clip ,scale=.055]{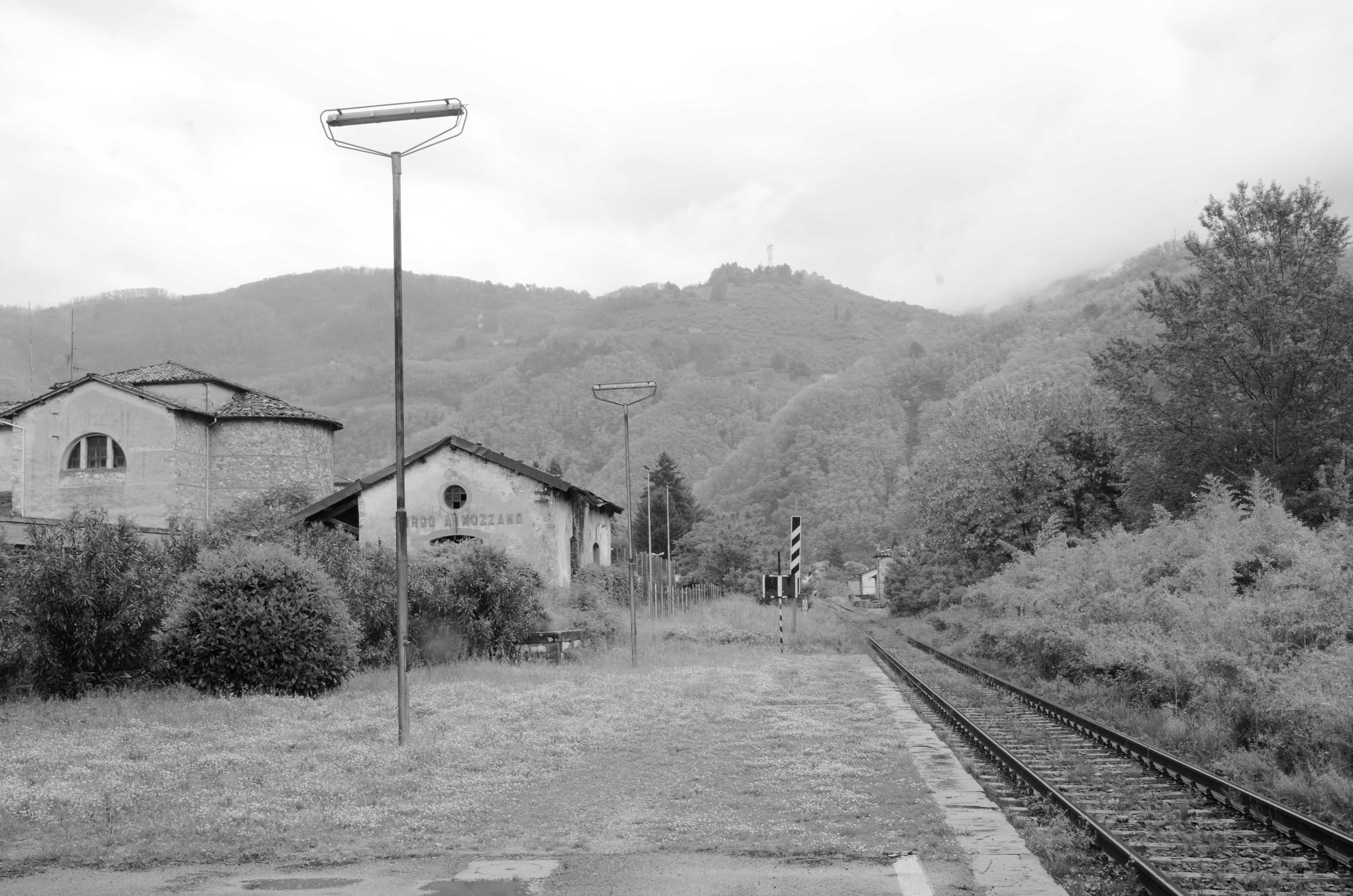}
\caption{Image compressed at a similarly low quality parameter to Figure \ref{fig:low_quality} but using our trained-at-95 table. No obvious gradation occurs.}
\label{fig:low_quality95}
\end{figure}

Previously we mentioned how our tables may introduce pixelation to compressed images, noticeable upon significant magnification. However, by accentuating the errors of our trained-at-95 table and the standard table, we can show how our table also increases the image fidelity on the unmagnified image and improves scaling. We already noted that in a side by side comparison our trained-at-95 table has little to no visual difference from the standard when the quality parameter is set to 95. Figure \ref{fig:low_quality} shows the same original image presented in Figure \ref{fig:95compiled_orig} compressed again by the standard table but with the quality parameter significantly lowered to accentuate errors.
The sky in this image shows clear gradation not in the original image. In fact, this kind of gradation of empty space is a common artifact of heavier jpeg compression. Compare this to the image presented in Figure \ref{fig:low_quality95}, which has been compressed using our trained-at-95 table, but using a similarly low quality parameter to Figure \ref{fig:low_quality}.
Figure \ref{fig:low_quality95} has no obvious gradation in the sky, nor obvious artifacts elsewhere. Despite this, the image is actually smaller than the one compressed by the standard table in Figure \ref{fig:low_quality}. 

\subsection{Timing}
One small concern for improving the quantization table is that somehow in the process of changing the quantization values, we would have slowed JPEG compression. However, as we ran over the 200 images in our evaluation set, none of our tables showed any statistically significant difference in timing from the standard table. Therefore, we are confident that these tables will not increase the overhead of JPEG compression.
\subsection{Understanding the Best Tables}
Examining Figure \ref{fig:heatmap}, we look at common trends throughout our proposed tables. The most noticeable decrease across the tables occurs in the DC coefficient, where in all but trained-at-35 it has been reduced from 16 to 8. In all training qualities except 95, the next 20 lowest frequencies have been significantly raised to counter-balance this change (frequencies are arranged diagonally), while higher frequencies have been decreased. This result matches the initial suggestions of Sherlock and Monroe that the standard table undervalues high frequencies. However, the case is flipped to an extent for trained-at-95. While the DC coefficient is still halved, higher frequency values have increased more significantly than the lower ones -- with the lowest frequencies being left essentially unchanged. The ubiquity in the decrease of the DC coefficient (indeed, this appeared in the majority of our runs across every annealing method) firmly supports the standard table undervaluing this coefficient, but the difference in trained-at-95 suggests that the relative importance of other frequencies may depend on the quality of the image. If this were the case, the quality scaling metric used by libjpg could be significantly improved to match the changing importance of these frequencies at different resolutions. 
\subsection{M-SSIM Runs}
In addition to our annealing on FSIM, we ran 400 annealing processes on quality 75 using the M-SSIM error metric. The results for these runs generate similar, but more erratic tables to those from FSIM. For instance, consider this table with over $10\%$ reduced error and $37\%$ improved compression:
\begin{equation*}
\setstacktabbedgap{3pt}
\text{M-SSIM} = \bracketMatrixstack{
13 & 17 & 21 & 39 & 45 & 81 & 65 & 75 \cr
24 & 32 & 0 & 19 & 45 & 113 & 110 & 54 \cr
19 & 29 & 50 & 57 & 75 & 105 & 115 & 70 \cr
18 & 40 & 27 & 45 & 89 & 109 & 119 & 61 \cr
28 & 47 & 56 & 115 & 131 & 127 & 128 & 69 \cr
56 & 35 & 77 & 118 & 161 & 171 & 87 & 109 \cr
87 & 125 & 105 & 138 & 163 & 167 & 182 & 137 \cr
117 & 146 & 156 & 180 & 138 & 136 & 147 & 118
}
\end{equation*}

The DC coefficient has been decreased and for the most part the lower frequencies have increased, but the table has odd behavior throughout--e.g. 0 at (2,3), or the 182 at (7,7). These are likely places where our annealing has exploited this less robust error metric. Regardless, the results show that our annealing technique is not limited to a single error metric. The only caveat is that both M-SSIM and FSIM operate under similar principles in their attempts to model the HVS. It would be interesting to attempt to expand our annealing to more distinct error metrics in future work

\section{Further Directions}
Our results on Butteraugli suggest that annealing over FSIM alone may not improve image fidelity in all aspects. Unfortunately, not only does there not exist a fully accurate error metric, but those that are powerful tend to be too slow to use for annealing. FSIM is an exception to this rule. However, it may be possible to get around this issue by using multiple error metrics. For instance, for every $x$ steps our annealing takes using the FSIM metric, 1 step could be run on Butteraugli to check that FSIM has not deviated too far from Butteraugli's model of the HVS. While this process would take more resources than our current annealing, it is still feasible to run over a cluster in a matter of weeks.
\section{Conclusion}
Through simulated annealing on FSIM we have produced quantization tables which significantly cut the size of JPEG files from $20-50\%$ on a variety of image qualities while improving errors in gradients on unmagnified images and quality scaling. This improvement sometimes comes at the trade-off of the resolution of the image upon magnification, and fixing this side-effect may be an interesting consideration for further work.

\section*{Acknowlegment}
Michael Mitzenmacher is supported in part by NSF grants CNS-1228598, CCF-1320231, CCF-1535795, and CCF-1563710.

\newpage

\section*{Appendix}
Please visit \url{http://www.eecs.harvard.edu/~michaelm/SimAnneal/PAPER/simulated-annealing-jpeg.pdf} for a copy of the paper including an appendix with detailed examples showing the annealing history and performance for more tables trained at varying qualities. The arXiv's size limitations preclude us from including all our results in this version of the paper. 
\end{document}